\documentclass[reprint,a4paper,preprintnumbers]{revtex4-1}

\usepackage[sumlimits,intlimits,namelimits]{amsmath} 
\usepackage{slashed}
\usepackage{amssymb}
\usepackage{bm}
\usepackage{revsymb}
\usepackage[english]{babel}
\usepackage{graphicx}
\usepackage{subfigure}
\usepackage{geometry}
\usepackage{hyperref}
\makeatletter
\def\fmslash{\@ifnextchar[{\fmsl@sh}{\fmsl@sh[0mu]}}
\def\fmsl@sh[#1]#2{%
  \mathchoice
    {\@fmsl@sh\displaystyle{#1}{#2}}%
    {\@fmsl@sh\textstyle{#1}{#2}}%
    {\@fmsl@sh\scriptstyle{#1}{#2}}%
    {\@fmsl@sh\scriptscriptstyle{#1}{#2}}}
\def\@fmsl@sh#1#2#3{\m@th\ooalign{$\hfil#1\mkern#2/\hfil$\crcr$#1#3$}}
\makeatother

\def\Ds{D^\ast}
\def\mB{m_B}
\def\mD{m_D}
\def\mDs{m_{D^\ast}}
\def\mDss{m_{D^{(\ast)}}}
\def\mb{m_b}
\def\mc{m_c}
\def\BtoDss{\bar B\to D^{(\ast)}\ell \bar\nu}

\def\btoDsln{\bar B \to D^\ast \ell \bar \nu}

\def\rds{r_\ast}
\def\Vcb{V_{cb}}

\def\ffG{\mathcal G}

\def\ffF{\mathcal F}

\def\ffA{\mathcal A}

\def\ffB{\mathcal B}

\def\Ops{\mathcal{O}^{\text{S}}}
\def\Opps{\mathcal{O}^{\text{PS}}}
\def\Opt{\mathcal{O}^{\text{T}}}
\def\Oppt{\mathcal{O}^{\text{PT}}}
\def\gs{g_{\text{S}}}
\def\gps{g_{\text{PS}}}
\def\gt{g_{\text{T}}}
\def\gpt{g_{\text{PT}}}
\def\Pp{P_R}
\def\Pm{P_L}
\def\cL{c_L}
\def\cR{c_R}
\def\gL{g_L}
\def\gR{g_R}
\def\dL{d_L}
\def\dR{d_R}
\def\cP{c_+}
\def\cM{c_-}
\def\dP{d_+}
\def\dM{d_-}
\def\gP{g_+}
\def\gM{g_-}
\def\bc{\bar c}
\begin{document}
\title{Limits on New Physics from exclusive \boldmath $B \to D^{(\ast)}\ell\, \bar\nu$ \unboldmath Decays}
\author{Sven \surname{Faller}}
\email[e-mail: ]{faller@tp1.physik.uni-siegen.de}
\author{Thomas \surname{Mannel}}
\email[e-mail: ]{mannel@tp1.physik.uni-siegen.de}
\affiliation{Theoretische Physik 1, Department Physik, Universit\"at Siegen, D-57068 Siegen, Germany}
\author{Sascha \surname{Turczyk}}
\email[e-mail: ]{ssturczyk@lbl.gov}
\affiliation{Theoretical Physics Group, Lawrence Berkeley National Laboratory, Berkeley, CA 94720\\
Theoretische Physik 1, Department Physik, Universit\"at Siegen, D-57068 Siegen, Germany}
\date{\today}
\preprint{SI-HEP-2010-06}

\begin{abstract}
\vspace{0.2cm}\noindent
We consider the exclusive decays $B \to D^{(\ast)}\,\ell\, \bar\nu$
and study the effect of non $V-A$ structures on the observables.
We extend the standard model hadronic current by additional right-handed vector
as well as left- and right-handed scalar
and tensor contributions and calculate the decay rates
including the perturbative corrections up to order $\alpha_s$.
Using the data of the exclusive semileptonic $b \to c $ decays
and recent calculations of the form factors at the non-recoil point
we discuss the constraints to the wrong helicity admixtures in the hadronic current.
\end{abstract}
\maketitle

\section{Introduction}
The $V-A$ structure of the  charged currents is considered a 
firmly established fact in weak interaction physics.
The evidence for $V-A$ is very strong for the leptonic couplings,
e.g. by the measurement of the Michel parameters of the muon decay.
However, this is not as clear for the hadronic currents due to our inability to 
perform a precise calculation of the hadronic matrix elements.
Consequently, it is hard to exclude admixtures of different helicities in hadronic charged currents. 

Over the last ten years, heavy-quark symmetries became a very useful tool in the calculation
of hadronic matrix elements involving heavy quarks.
In particular, they may help to perform the analogue of a Michel parameter analysis for the hadronic charged currents.
Making use of the detailed data from the flavor factories the semileptonic heavy quark decays 
may serve as a sensitive test for  possible ``wrong-helicity'' contributions.  

In a recent analysis we extracted limits on a right handed admixture from the wealth of data 
on inclusive semileptonic decays \cite{Feger:2010qc}.
Despite the large amount of data for the inclusive semileptonic decays and the precise theoretical tools,
it turns out that the exclusive  decays $B \to D^{(\ast)}\ell \bar\nu$
can be much more sensitive to wrong helicity admixtures than the inclusive decays. 

In this paper we expand on this idea,
including also scalar and tensor components for the hadronic current.
The method we propose can be most easily explained in the Isgur-Wise limit for the decays $B \to D^{(\ast)}\ell \bar\nu$,
where only a single form factor appears,
of which  the normalization is known.
Starting from this limit, the corrections may be considered 
and the method can be systematically improved. 

Beyond the Isgur-Wise limit a large number of form factors appears,
most of which are not well studied.
However, based on the detailed analysis of the vector and axial-vector form factors through lattice and 
QCD sum rule calculations one may perform a stringent test at least for a possible right-handed admixture.  

The paper is organized as follows.
In the next section we study a possible new physics (NP) contribution
in the Isgur-Wise limit to demonstrate how the proposed method works.
After that we will calculate the corresponding radiative corrections,
followed by a section dealing with the calculation of bounds on right-handed
admixtures regarding the contemporary experimental results as well as lattice and non-lattice calculations
to be able to provide a comparison to the standard model (SM) results.
Finally we discuss our results and give a prospect into possible additional measurements.

\section{New Physics contributions in the Isgur-Wise limit}\label{New Physics contributions in the IW limit}
As has been pointed out above,
the $V-A$ structure of the leptonic current is well established and
hence we do not modify this current.
However, the hadronic current may contain a contribution from ``new physics''
and hence the effective Hamiltonian to be considered is 
\begin{equation}
H_{\text{eff}}= \frac{4 G_F \Vcb}{\sqrt{2}} \ J_{h,\mu}  (\bar e \gamma^\mu \Pm \nu_e) , 
\end{equation}
where $J_{h,\mu}$ is the generalized hadronic current
and $P_{L,R}= (\openone  \mp \gamma_5)/2$ is the projector of negative/positive chirality.
The modifications in the hadronic current can be considered on the basis of an effective field theory approach
and one obtains (see eg. \cite{Dassinger:2008as})
\begin{eqnarray} 
 J_{h,\mu} &= & \cL \bc  \gamma_\mu \Pm b + \cR \bc  \gamma_\mu \Pp b \nonumber \\ 
  &+&  \gL \bc \left( i \tensor D_\mu \right) \Pm b 
 + \gR \bc \left( i \tensor D_\mu \right) \Pp b \nonumber \\ 
 &+& \dL  i \partial^\nu (\bc\, i \sigma_{\mu\nu} \Pm b)
 + \dR\, i \partial^\nu (\bc\, i \sigma_{\mu\nu} \Pp b) ,\label{hadcur}
\end{eqnarray}
where $D_\mu$ is the QCD covariant derivative and
$ f \tensor D_\mu g = f(D_\mu g) - (D_\mu f)g$
represents the left and right derivatives.
Note that the first line in (\ref{hadcur})
corresponds to dimension-3 operators with dimensionless couplings $\cL$ and $\cR$,
while the second and third lines are dimension-4 operators with dimensionful couplings $g_{L/R}$ and 
$d_{L/R}$. There 
are two other dimension-4 operator of the form $(\mb + \mc) \bc \gamma_\mu b$ and 
$(m_b - m_c) \bc \gamma_\mu \gamma_5 b$, which are related by the Gordon identities
\begin{eqnarray}
 (\mb + \mc) \bc \,\gamma_\mu b  
 &=& \bar c \, i \tensor D_\mu b \nonumber\\
  &&+ i \partial^\nu \left( \bar c (- i \sigma_{\mu \nu}) b \right)  , \label{GIV}\\
 - (\mb - \mc) \bar c \,\gamma_\mu\gamma^5 b  
 &=& \bc \, i \tensor D_\mu \gamma^5b \nonumber \\
  &&+ i \partial^\nu \left( \bc (- i \sigma_{\mu \nu}) \gamma^5 b
 \right)  . \label{GIPV}
\end{eqnarray}  
Hence these operators are redundant
and do not need to be considered seperately as it has been done in \cite{Dassinger:2008as}.
Likewise, the pseudotensor is not independent of the tensor due to the relation 
\begin{equation}
 \bc \sigma_{\mu \nu} \gamma_5 b = -\frac i2 \epsilon_{\mu \nu \alpha \beta}  \bc \sigma^{\alpha \beta} b ,
\end{equation}
but it is convenient to keep this operator explicitely. 

From the effective field theory analysis performed in \cite{Dassinger:2008as}
all these operators originate from dimension-6 operators parametrizing physics beyond the standard model.
From this one obtains
\begin{equation}
 \begin{split}
\cL \sim 1_{\text{SM}} &+  {\mathcal O} \left(\frac{v^2}{\Lambda^2} \right), \
\cR \sim   {\mathcal O} \left(\frac{v^2}{\Lambda^2} \right) , \\
&g_{L/R},\, d_{L/R} \sim  \frac{1}{v}  {\mathcal O} \left(\frac{v^2}{\Lambda^2} \right) ,
\end{split}
\end{equation} 
where $\Lambda$ is the scale of ``new physics''
and $v$ is the electroweak vacuum expectation value. 

We shall use this hadronic current to evaluate the semileptonic widths for $B \to D^{(*)} \ell \bar{\nu}$.
We first study the Isgur-Wise limit where the relevant kinematic quantity is the product of the four 
velocities of the initial and final state hadrons
\begin{equation}\label{wdef}
 w = v \cdot v' = \frac{\mB^2 + \mDss^2-q^2}{2 \mB \mDss} \ ,
\end{equation}
where $q^2 \equiv (p - p')^2$,
and $p$ and $p'$ refer to the four-momenta
of the $B$ and $D^{(\ast)}$ mesons, respectively.
The differential exclusive decay rates for the $D$ and $\Ds$ mesons
can as usual be expressed in terms of the hadronic form factors $\ffG (w)$
and $\ffF(w)$, respectively.
Thus the differential decay rates read
\begin{eqnarray}
    \frac{d\Gamma}{dw} &=& G_0 (w) |\Vcb|^2\,\frac{w-1}{w+1}(1+r)^2|\ffG (w)|^2,\label{diffdecrateI}\\
    \frac{d\Gamma^{*}}{dw} &=& G_0^{*} (w) |V_{cb}|^2\,|\ffF (w)|^2,\label{diffdecrateII}
\end{eqnarray}
where we defined the factors
\begin{equation}
G_0^{(\ast)} (w) =  \frac{G_F^2m_B^5 }{48\pi^3}  r_{(\ast )}^3 \sqrt{w^2 -1} (w+1)^2 \ ,
\end{equation}
containing the kinematical and normalization coefficients to streamline the notation and 
$r_{(\ast)}= \mDss/m_B$.
The form factors $\ffF (w)$ and $\ffG (w)$
are related to matrix elements of the hadronic current, 
which in our case also contains the new physics effects represented
by the couplings $\cR$, $d_{L/R}$ and $g_{L/R}$. 

The Isgur-Wise limit is taken by letting $\mb,\mc  \to \infty $
with $\mc / \mb \sim {\mathcal O}(1)$.
In this case both the charm and the bottom quark in the hadronic current
have to be replaced by static quarks $h_{v',c}$ and $h_{v,b}$;
to leading order,
the hadronic current matches onto 
\begin{eqnarray}
 J_{h,\mu} &=& \cL \bar{h}_{v',c} \gamma_\mu \Pm h_{v,b} + \cR \bar{h}_{v',c} \gamma_\mu \Pp h_{v,b}\nonumber\\
  &&+ \gL (\mb v_\mu + \mc v'_\mu) \bar{h}_{v',c} \Pm h_{v,b}\nonumber\\
  &&+ \gR (m_b v_\mu + m_c v'_\mu) \bar{h}_{v',c} \Pp h_{v,b} \nonumber\\
  &&+ \dL  (\mb v^\nu - \mc v^{\prime \nu}) (\bar{h}_{v',c}i \sigma_{\mu\nu} \Pm h_{v,b})\nonumber\\
  &&+ \dR  (\mb v^\nu - \mc v^{\prime \nu}) (\bar{h}_{v',c} i \sigma_{\mu\nu} \Pp h_{v,b})  ,\label{hadcurII}
\end{eqnarray}
and all the relevant matrix element
can be expressed in terms of the Isgur-Wise function $\xi (w)$ 
\cite{Isgur:1989vq,Isgur:1989ed,Shifman:1987rj},
which is normalized to $\xi(1) \equiv 1$ at zero recoil
\begin{eqnarray}
\frac{\langle D(v')| \bar{h}_{v',c} h_{v,b} | B(v)\rangle}{\sqrt{\mB\mD}} &=& (1 + w) \xi(w)\label{BtoD} , \\
\frac{\langle D(v')| \bar{h}_{v',c} \gamma^\mu h_{v,b} | B(v)\rangle}{\sqrt{\mB\mD}} &=& (v+v')^\mu \xi (w)  ,  \\
\frac{\langle D(v')| \bar{h}_{v',c} \sigma^{\mu \nu}  h_{v,b} | B(v)\rangle}{\sqrt{\mB \mD}}&&\nonumber\\
      =  i (v_\mu'v_\nu  &-& v_\nu' v_\mu) \xi(w) , 
\end{eqnarray}
for the $B \to D\,\ell\,\bar\nu$ decay and 
\begin{widetext}
\begin{eqnarray}
\frac{\langle \Ds(v',\epsilon)|\bar{h}_{v',c} \gamma_5 h_{v,b}| B(v)\rangle}{\sqrt{\mB \mDs}} 
&=&  (\epsilon^\ast \cdot v) \xi (w) \ ,  \\
\frac{\langle \Ds (v',\epsilon)|\bar{h}_{v',c} \gamma^\mu h_{v,b}|  B(v)\rangle}{\sqrt{\mB\mDs}}
&=& i   \varepsilon^{\mu\nu\alpha\beta}\epsilon^\ast_\nu v_\alpha'v_\beta \xi(w)\ , \\
\frac{\langle \Ds (v',\epsilon)|\bar{h}_{v',c}  \gamma^\mu \gamma_5 h_{v,b}| B(v)\rangle}{\sqrt{\mB\mDs}} 
&=&  \left[ (1+w) \epsilon^{\ast\mu} + v^\mu (v \cdot \epsilon^\ast) \right] \xi(w)\ ,  \\  
\frac{\langle \Ds(v',\epsilon)|\bar{h}_{v',c}  \sigma^{\mu\nu} h_{v,b}|B(v)\rangle}{\sqrt{\mB \mDs}} 
&=&\ \varepsilon^{\mu\nu\kappa\tau}  \epsilon^\ast_\kappa (v'+v)_\tau \xi (w) \ .  \label{BtoDst}
\end{eqnarray}
\end{widetext}
for the $\bar B \to D^\ast\ell\bar\nu$ decay.
The decay rate for the semileptonic $\BtoDss$ decays
in the Isgur-Wise limit can thus be expressed solely
in terms of the Isgur Wise function, even in the presence of ``new physics''.
The form factors become 
\begin{eqnarray}
    |\ffG(w)|^2 &=& \frac{w+1}{w-1}(1+r)^{-2}\, \ffA(w) |\xi(w)|^2\ , \\
    |\ffF(w)|^2 &=& \left(\ffB^T(w) + \ffB^L(w)\right) |\xi(w)|^2,
\end{eqnarray}
where we have separated the rate for $B \to D^\ast$
into the contributions for longitudinally and transversely polarized $D^\ast$ mesons.
Hence we end up with
\begin{eqnarray}
\frac{d\Gamma^{B \to D \ell \bar{\nu}}}{dw}  &=& G_0 (w) 
 |\Vcb|^2  \ffA (w)  | \xi (w) |^2\ , \\
\frac{d\Gamma^{B \to D^\ast_T \ell \bar{\nu}}}{dw}   &=&G_0^\ast (w) 
 |\Vcb|^2 \ffB^T (w)  | \xi (w) |^2\ , \\
\frac{d\Gamma^{B \to D^\ast_L \ell \bar{\nu}}}{dw}   &=&G_0^\ast (w) 
 |\Vcb|^2 \ffB^L (w)  | \xi (w) |^2 \ ,\\
 \frac{d\Gamma^{B \to D^\ast  \ell \bar{\nu}}}{dw}   &=& G_0^\ast (w) 
 |\Vcb|^2 \nonumber\\
  && \times\left[ \ffB^T (w) + \ffB^L (w) \right] | \xi (w) |^2 \ , 
\end{eqnarray} 
where the coefficient functions $\ffA (w)$, $\ffB^T$
and $\ffB^L$ contain the dependence on the new physics couplings. 
\begin{eqnarray}  
\ffA (w) &=&\frac{w-1}{w+1}\bigl\lbrack \cP (1+r) -\mB  \dP (r^2-2rw+1) \nonumber\\
  &&+2 \mB r \gP  (w+1)\bigr\rbrack^2\ , \label{NP-BtoDpar} \\
\ffB^T (w) &=& 2\lbrack1- 2 \rds w + (\rds)^2 \rbrack \Bigl 
\lbrace \lbrack \cM +\dM \mB (\rds-1)\rbrack^2 \nonumber\\
&&+\frac{w-1}{w+1}\bigl\lbrack\cP + \dP \mB (\rds+1)\bigr\rbrack^2 \Bigr 
\rbrace\ , \label{NP-BtoDsT} \\
\ffB^L (w) &=& \Bigl\lbrace \cM (\rds-1)+2 \gM \mB \rds (w-1) \nonumber\\
&&+\dM \mB  \lbrack(\rds)^2 -2  \rds  w +1\rbrack\Bigr\rbrace^2\ , \label{NP-BtoDsL}
\end{eqnarray}
where we define the combinations of coupling constants 
\begin{equation}
\begin{split}
c_\pm \equiv (\cL &\pm \cR), \ d_\pm \equiv (\dL\pm \dR) , \\
&  g_\pm \equiv (\gL \pm \gR).
\end{split}
\end{equation}
The expressions of the standard model are retrieved by setting $c_\pm = 1$
and all other couplings zero,
and we recover the standard model case as 
\begin{eqnarray}
\ffA_{\text{SM}}   (w) &=&\frac{w-1}{w+1} (1+r)^2\ , \\
\ffB^T_{\text{SM}}  (w) &=&   \frac{4w}{w+1} \lbrack1- 2 \rds w + \rds^2 \rbrack\ ,  \\
\ffB^L_{\text{SM}} (w) &=&   (\rds-1)^2 \ .
\end{eqnarray}
In the context of the extraction of $|V_{cb}|$ from exclusive decays
the measured $w$ spectrum is extrapolated to  $w =1$
and hence one studies the observables 
\begin{eqnarray} \label{Ms} 
M (w) &=&  \frac{d\Gamma^{B \to D \ell \bar{\nu}}}{dw}  \, \frac{1}{G_0} 
\, \frac{1}{\ffA_{\text{SM}} (w)}\ , \\ 
M^\ast_T(w)  &=&  \frac{d\Gamma^{B \to D^\ast_T \ell \bar{\nu}}}{dw}  \, \frac{1}{G_0^\ast}  
\frac{1}{\ffB^T_{\text{SM}} (w)}\ , \\
M^\ast_L (w) &=&  \frac{d\Gamma^{B \to D^\ast_L \ell \bar{\nu}}}{dw}  \, \frac{1}{G_0^\ast}   
\frac{1}{ \ffB^L_{\text{SM}} (w)}\ ,  \\ 
M^\ast (w) &=&  \frac{d\Gamma^{B \to D^\ast  \ell \bar{\nu}}}{dw}  \, \frac{1}{G_0^\ast}   
\frac{1}{\ffB^L_{\text{SM}}  (w) + \ffB^T_{\text{SM}} (w)}  .
 \label{Mss}
\end{eqnarray} 
These observables become in the standard model  
in all three cases just $| V_{cb}| ^2 | \xi (w) |^2$.
Extrapolating $M(w)$ and/or  $M^\ast(w)$ to the zero-recoil point $w=1$
and making use of the normalization of the Isgur Wise function allows us to extract $|V_{cb}|$ model independently.
Expanding around $w=1$ and using 
\begin{equation}\label{ExpIWF}
\xi (w) =  \xi (1) \left[ 1- \rho_{\text{\text{IW}}}^2 (w-1) + \dots \right] \ ,
\end{equation}
we may also obtain information on the slope of the Isgur-Wise function by performing the 
corresponding expansion of the expressions \eqref{Ms}--\eqref{Mss}.
In the standard model we obtain: 
\begin{eqnarray}
 \rho_{\text{\text{IW}}}^2 &=& - \frac{1}{2 M(1)}\left.  \frac{\partial M(w)}{\partial w} \right|_{w=1} \nonumber\\
    &=& - \frac{1}{2M_T^\ast(1)} \left.  \frac{\partial M_T^\ast(w)}{\partial w} \right|_{w=1} \nonumber  \\
    &=& -   \frac{1}{ 2M_L^\ast(1)} \left.  \frac{\partial M_L^\ast(w)}{\partial w} \right|_{w=1}\nonumber\\ 
    &=& - \frac{1}{2  M^\ast(1)} \left.  \frac{\partial M^\ast(w)}{\partial w} \right|_{w=1}\ .
\end{eqnarray} 
However, the presence of the new physics contributions will change these relation
and allows us to re-interpret the measured observables in terms of a possible ``new physics'' contribution.  
In fact,
inserting the general expressions into \eqref{Ms}--\eqref{Mss},
we get 
\begin{eqnarray} \label{MsNP} 
M (w) &=&  \frac{ \ffA (w) }{ \ffA_{\text{SM}} (w)} | V_{cb}| ^2 | \xi (w) |^2\ , \\ 
M^\ast_T(w)  &=&  \frac{ \ffB_T (w)}{\ffB^T_{\text{SM}} (w)} | V_{cb}| ^2 | \xi (w) |^2 \ ,  \\
M^\ast_L (w) &=&   \frac{ \ffB_L (w)}{ \ffB^L_{\text{SM}} (w)} | V_{cb}| ^2 | \xi (w) |^2 \ , \\ 
M^\ast (w) &=&   \frac{ \ffB_T (w)+ \ffB_L (w) }{\ffB^T_{\text{SM}} (w) + \ffB^L_{\text{SM}} (w)} | V_{cb}| ^2 | \xi (w) |^2 \ .
 \label{MssNP}
\end{eqnarray} 
At the zero-recoil point $w=1$ this becomes
\begin{eqnarray}
 M (1) &=&  | V_{cb}| ^2 | \xi (1) |^2 \biggr[c_+ - m_B d_+ \frac{(r-1)^2}{r+1} \nonumber \\
  && + 4 m_B g_+ \frac{r}{r+1} \biggr]^2 \ , \\ 
 M_T^\ast (1) &=& | V_{cb}| ^2 | \xi (1) |^2 \left[c_- + (r^\ast-1) m_B d_- \right]^2\nonumber\\
  &=& M_L^\ast (1) = M^\ast (1)\ ,
\end{eqnarray} 
leading to
\begin{eqnarray} 
&-& \frac{1}{2 M(1)}\left.  \frac{\partial M(w)}{\partial w} \right|_{w=1} =  \rho_{\text{IW}}^2  \nonumber\\
&&  + \frac{2 r m_B (\dP + \gP)}{\mB \dP (r-1)^2 - (r+1) \cP - 4 r \mB \gP}  ,  \\ 
&-& \frac{1}{2 M_T^\ast(1)}\left.  \frac{\partial M_T^\ast(w)}{\partial w} \right|_{w=1} \nonumber\\
&&=  \rho_{\text{IW}}^2 + \frac{1}{4} \left[1 - 
 \left(\frac{\cP +(\rds + 1) \mB \dP}{\cM + (\rds - 1)\mB \dM}\right)^2 \right]  , \\ 
&-& \frac{1}{2 M_L^\ast(1)}\left.  \frac{\partial M_L^\ast(w)}{\partial w} \right|_{w=1} \nonumber\\
&&= \rho_{\text{IW}}^2 - \frac{2 \rds m_B (\dM - \dM)}{(\rds -1)(\cM + (\rds-1)\dM)}  , \\ 
&-& \frac{1}{2 M^\ast(1)}\left.  \frac{\partial M^\ast(w)}{\partial w} \right|_{w=1} = \rho_{\text{IW}}^2 \nonumber\\
&&+ \frac{1}{6}\biggl\lbrace 1- \left(\frac{\cP+\dP\mB(\rds+1)}{\cM+\dM\mB (\rds-1)}\right)^2  \nonumber\\
&& + \frac{4\mB\rds (\dM - \gM)}{(\rds-1)\lbrack\cM+\dM \mB (\rds-1)\rbrack}\biggr\rbrace\ ,
\end{eqnarray}
when we calculate the slopes.
Note that the current analyses are performed for the total
$B \to D^\ast\,\ell\,\bar\nu$ rate
without the decomposition into longitudinal and transversal polarizations. 

\section{Radiative Corrections}
Up to this point we have been relying on the Isgur-Wise limit.
However, from lattice as well as from QCD sum rule studies we know
that the relations obtained in the infinite mass limit may have corrections of 
the order of ten percent,
making a sensitive test on the basis of the formulae in the last section impossible. 

The first class of corrections are the perturbative QCD corrections
which break the Isgur Wise symmetry. 
The dimension-3 operators, i.e.
the vector $\bar{c} \gamma_\mu b$  and the axial vector 
$\bar{c} \gamma_\mu \gamma_5 b$
are both conserved in the massless lepton limit
and hence do not have an anomalous dimension.
The dimension-4 operators   
\begin{align}
  \Ops &= \bar c i \tensor D_\mu b\ , \\
  \Opps &= \bar c i \tensor D_\mu \gamma^5 b\ , \\
  \Opt&= i \partial^\nu \bar c (- i \sigma_{\mu \nu}) b\ , \\
  \Oppt&= i \partial^\nu \bar c (- i \sigma_{\mu \nu}) \gamma^5 b \ ,
\end{align}
mix under renomalization and also have anomalous dimensions \cite{Dassinger:2008as}.
Note that $\Opt$ and $\Oppt$ are not independent due to the relation 
$\sigma_{\mu \nu} \gamma_5 = \frac{i}{2} \epsilon_{\mu \nu \alpha \beta} \sigma^{\alpha \beta}$. However, 
for our purposes it is useful to keep both operators.

Gathering the four operators and the coupling constants in columns  
\begin{equation}
\begin{split}
&\vec{{\mathcal O}} (\mu) =  \begin{pmatrix} 
    \Ops (\mu) \\
    \Opps (\mu) \\
    \Opt (\mu) \\
    \Oppt (\mu) 
 \end{pmatrix} \ , \\
&\vec{G} (\mu) =  \begin{pmatrix} 
    \gs (\mu) \\
    \gps (\mu) \\
    \gt (\mu) \\
    \gpt (\mu) 
 \end{pmatrix}  = 
 \begin{pmatrix} 
    \gL(\mu) + \gR (\mu)  \\
    \gR(\mu) - \gL (\mu)  \\
    \dL(\mu) + \dR (\mu)  \\
    \dR(\mu) - \dL (\mu)  
 \end{pmatrix}\ , 
\end{split}
\end{equation} 
we can write the hadronic current \eqref{hadcur} as  
\begin{equation}
J_{h,\mu} = c_L \, \bc  \gamma_\mu \Pm b + c_R \, \bc  \gamma_\mu \Pp b + \vec{G} (\mu) \cdot \vec{{\cal O}}(\mu) \ .
\end{equation} 

The one-loop anomalous dimension matrix can be calculated and becomes
\begin{equation}
  \gamma = \frac{\alpha_s}{4\pi} C_F 
\begin{pmatrix}
 6 & 0 & 4 & 0 \\
 0 & 6 & 0 & 4 \\
 0 & 0 & 2 & 0 \\
 0 & 0 & 0 & 2
\end{pmatrix}
\end{equation}
and the solution of the renormalization group equation can be written as 
\begin{widetext}
\begin{equation}\label{wilsons_full}
    \begin{pmatrix} 
	\gs (\mu) \\
	\gps (\mu) \\
	\gt (\mu) \\
	\gpt (\mu) 
    \end{pmatrix} 
    = 
\begin{pmatrix}
 \gs (\Lambda ) \left(\frac{\alpha _s(\Lambda )}{\alpha _s(\mu )}\right){}^{\frac{3 C_F}{\beta _0}} \\
 \gps (\Lambda ) \left(\frac{\alpha _s(\Lambda )}{\alpha _s(\mu )}\right){}^{\frac{3 C_F}{\beta _0}} \\
 \left(\gt (\Lambda )+ \gs (\Lambda ) 
  \left(\left(\frac{\alpha _s(\Lambda )}{\alpha _s(\mu )}\right){}^{\frac{2 C_F}{\beta _0}}-1\right)\right) \left(\frac{\alpha _s(\Lambda )}{\alpha _s(\mu )}\right){}^{\frac{C_F}{\beta _0}}
   \\
 \left( \gpt (\Lambda )+ \gs (\Lambda ) \left(\left(\frac{\alpha _s(\Lambda )}{\alpha _s(\mu )}\right){}^{\frac{2 C_F}{\beta _0}}-1\right)\right) \left(\frac{\alpha _s(\Lambda )}{\alpha _s(\mu
   )}\right){}^{\frac{C_F}{\beta _0}}
\end{pmatrix}\ .
\end{equation} 
\end{widetext}
Note that his result is compatible with the Gordon Identities \eqref{GIV} and \eqref{GIPV}.
The left hand side of these identities only have the anomalous dimension of the masses 
\begin{equation}
\mu \frac{\partial}{\partial \mu} m (\mu) = m (\mu)  \gamma_m (\alpha_s (\mu)) 
\end{equation} 
with $\gamma_m = 6 \alpha_s C_F/(4 \pi)$
and hence we may check 
\begin{eqnarray} 
\gamma_m   (m_b &+& m_c) \bar c \,\gamma_\mu b = \gamma \cdot \begin{pmatrix} 
    \Ops (\mu) \\
    0 \\
    \Opt(\mu) \\
    0 
 \end{pmatrix} \nonumber\\
&=&  \frac{6 \alpha_s}{4\pi} C_F \left[ \bar c \, i \tensor D_\mu b + i \partial^\nu \left( \bar c (- i \sigma_{\mu \nu}) b \right) \right]
\end{eqnarray} 
and likewise for the axial vector. 

The $\mu$ dependence of the operators
has to be cancelled by the corresponding dependence of the matrix elements of the operators.
Here we focus on the matrix elements at the specific kinematic point $v = v'$ or $w =1$.
At this point,
all possible Dirac structures can be expressed in terms of the four matrices \cite{Mannel:1994kv} 
\begin{equation}
\openone = \frac{1}{2} (1 + \fmslash{v}) ,  \ s_\mu = \frac{1}{4} (1 + \fmslash{v}) \gamma_\mu \gamma_5 (1 + \fmslash{v})
\end{equation} 
with $v\cdot s = 0$.
To this end, we can write 
\begin{eqnarray} 
\bigl\langle c(p_c &=& m_c v) \bigl| \bar c  \gamma_\mu b \bigr| b (p_b = m_b v) \bigr\rangle \nonumber\\
&&  = \eta_{\text{V}} \, v_\mu  \bar{u}_c (v)  u_b (v)\ ,  \\ 
\bigl\langle c(p_c &=& m_c v) \bigl| \bar c  \gamma_\mu \gamma_5 b \bigr| b (p_b = m_b v) \bigr\rangle\nonumber\\
 &&= \eta_{\text{A}}  \, \bar{u}_c (v) s_\mu u_b (v)  \ ,
\end{eqnarray} 
while out of the dimension-4 operators we have a priori four additional matrix elements.
The matrix element of $\Opt$ vanishes at non-recoil,
while the one of $\Oppt$ does not.
We choose to use $\Ops$ and $\Opps$,
which have nonvanishing matrix elements at $w = 1$,
\begin{eqnarray} 
&&\left\langle c(p_c = m_c v) \left| \bar c \, i \tensor D_\mu b \right| b (p_b = m_b v) \right\rangle \biggr|_\mu   \nonumber\\
&=& v_\mu  \bar{u}_c (v)  u_b (v) \, (m_b + m_c) \eta_\text{S} (\mu)  \ ,  \\
&&\left\langle c(p_c = m_c v) \left| \bar c \, i \tensor D_\mu \gamma^5 b \right| b(p_b = m_b v) \right\rangle\biggr|_\mu \nonumber\\
&=& \bar{u}_c (v) s_\mu u_b(v) \, (m_b - m_c) \eta_\text{PS}(\mu) \, . 
\end{eqnarray}
The matrix element of the nonvanishing pseudotensor operator $\Oppt$
can according to the Gordon identity \eqref{GIPV} be expressed through
the axial vector times $(m_b-m_c)$ masses in the corresponding $\overline{\text{MS}}$ scheme.
Thus it can be expressed by
\begin{eqnarray}
  &&\left\langle c(p_c = m_c v) \left| i \partial^\nu \left( \bc (- i \sigma_{\mu \nu}) \gamma^5 b \right)\right| b (p_b = m_b v) \right\rangle \biggr|_\mu\nonumber\\
  &=& \bar{u}_c (v) s_\mu u_b(v) \, (m_b - m_c) \eta_\text{PT}(\mu) \\
  &=&  - (\overline{m}_b-\overline{m}_c) (\eta_{\text{A}} +\eta_\text{PS}) \, \bar{u}_c (v) s_\mu u_b (v) \,.\label{ptme}
\end{eqnarray}
\begin{widetext}
The matrix elements of the vector and axial-vector currents
are known at two-loops in the full phase space, 
while the matrix elements of the dimension-4 currents
are calculated here only at the non-recoil point 
$w = 1$.  The result is 
\begin{eqnarray}
\eta_{\text{A}} &=& 1+ \frac{\alpha_s}{4\pi}C_F \left[-8+3\frac{\mb+\mc}{\mb-\mc}\log\frac{\mb}{\mc} \right]\ ,  \\
\eta_{\text{V}} &=&   1+\frac{\alpha_s}{4\pi} C_F \left[-6+3\frac{\mb+\mc}{\mb-\mc}\log\frac{\mb}{\mc} \right]\ , \\
\eta_{\text{S}} (\mu)  &=& 1 +  \frac{\alpha_s }{4\pi}C_F 
\Bigg[ - 3 \log\left(\frac{\mu^2}{\mb \mc} \right)  
 + 6 \frac{\mb^2+\mc^2 }{\mb^2-\mc^2}\log \frac{\mb}{\mc}-10 \Bigg] \ , \\
\eta_{\text{PS}}(\mu) &=& \frac{\alpha_s}{4\pi}C_F \left[ -2 \log\left(\frac{\mu^2}{\mb \mc}\right) + 2 \frac{\mb + \mc}{\mb - \mc} \log\frac{\mb}{\mc} -4\right]\ , \\
 \eta_{\text{PT}}(\mu)&=& 1+ \frac{\alpha_s}{4\pi} C_F \left[ - \log\left(\frac{\mu^2}{\mb \mc}\right) + 4 \frac{\mb+\mc}{\mb-\mc}\log\frac{\mb}{\mc}-8 \right]\ . 
\end{eqnarray} 
\end{widetext}
Note that the difference in $|\eta_{\text{PS}}+\eta_{\text{A}}| \neq |\eta_{\text{PT}}|$
in \eqref{ptme} is due to the $\alpha_s$ pieces of the $\overline{\text{MS}}$ masses.
It is easy to check that the $\mu$ dependence cancels between the matrix elements
and the Wilson coefficients in the order $\alpha_s$.
However, the renormalization group flow resumms potentially large logarithms of the form
$(\alpha_s / \pi)^n  \ln^n (\Lambda^2 / \mu^2)$
and hence there will be a residual $\mu$ dependence. 
Looking at the structure of the matrix element coefficients $\eta_X(\mu)$,
$X\in\{\text{A, V, S, PS, PT}\}$, of the dimension four operators,
a natural choice is $\mu_0^2 = m_b m_c$ and hence we insert this scale for our numerical study.
This includes that the couplings generated at some high scale $\Lambda$
have to be evolved down to this small scale.  

Numerically we obtain using the values $m_b = 4.2$~GeV,
$m_c = 1.3$~ GeV (and thus $\mu_0 \approx 2.34$~GeV), and $C_F = 4/3$ as well as $\alpha_s(\mu_0) \approx 0.281 $ \cite{Bethke:2009jm} for $N_f = 3$
\begin{subequations}
\begin{alignat}{2}
    \eta_{\text{V}}  &\approx 1+0.0713\, \alpha_s &\, &\approx 1.02 \ , \label{etaV}\\
    \eta_{\text{A}} &\approx 1-0.1409\, \alpha_s &\, &\approx 0.96 \ , \label{etaA}\\
    \eta_\text{S} (\mu_0) &\approx 1 - 0.1562\, \alpha_s &\, &\approx 0.96 \ ,\label{etaS}\\
    \eta_\text{PS} (\mu_0) &\approx 0+0.0476\, \alpha_s &\, &\approx 0.01 \ , \label{etaPS}\\
    \eta_\text{PT} (\mu_0) &\approx 1 + 0.0951\, \alpha_s &\, &\approx 1.03 \ . \label{etaPT}
\end{alignat}
\end{subequations}
Comparing the results for the vector and axial vector coefficients
Eqs. \eqref{etaV} and \eqref{etaA}, respectively,
with the  form factors  $\ffG(1)$ and $\ffF(1)$
from lattice or non-lattice calculations,
as discussed in more detail in the next section,
the results from our calculation can be 
assumed as a first approximation for these form factors.
Following the same line the results
Eqs. \eqref{etaS}--\eqref{etaPT} can 
be considered as first approximations for
scalar, pseudoscalar and pseudotensor form factors values at the non-recoil point.

\section{Constraints on right-handed admixtures}
In this section we shall discuss the bounds on possible admixtures to the standard model current.
In contrast to section \ref{New Physics contributions in the IW limit}
we will perform the analysis not in the Isgur-Wise Limit, and hence we have to deal with the 
form factor values at zero recoil. From lattice simulations as well as from QCD sum rules 
we know the normalizations for the vector- and the axial-vector form factors, and hence we can 
- off the Isgur Wise limit - only 
study a possible right-handed admixture to the weak hadronic currents,
which shows up to be the best candidate for sizable contributions \cite{Dassinger:2007pj}.
To be able to extract the strength of the right-handed admixture in the weak currents of exclusive decays,
we start from the  exclusive differential decay rates \eqref{diffdecrateI} and \eqref{diffdecrateII}
of the $D$ and $\Ds$ mesons, respectively.
All information about the right-handed admixture is contained in the form factors $\mathcal F(w)$ and $\mathcal G(w)$.
Like for the Isgur-Wise function $\xi(w)$ we may extrapolate the form factors to the point $w=1$
and perform an expansion around this point to express the value for any other $w$
by a small correction of order $\Lambda_{\text{QCD}}/m_Q$,
where we use $m_Q$ generically for $m_b$ or $m_c$ respectively.
The form factors $\ffF(w)$ and $\ffG(w)$ can be expanded as
\begin{eqnarray}
    \ffF(w) &=& \ffF(1) \nonumber \\
    &\times& \left[ 1 - \rho_\ast^2 (w-1) + c_\ast(w-1)^2 + \dots \right] \label{Fexpansion} ,\\
    \ffG(w) &=& \ffG(1) \nonumber \\
  &\times& \left[ 1 - \rho^2(w-1) + c(w-1)^2 + \dots\right]  ,  \label{Gexpansion} 
\end{eqnarray}
where the slopes
\begin{eqnarray}
    \rho_\ast^2 &=&  -\frac{1}{\ffF(1)} \frac{\partial\ffF(w)}{\partial w} \bigg|_{w=1} \ , \\
    \rho^2 &=&  -\frac{1}{\ffG(1)} \frac{\partial\ffG(w)}{\partial w} \bigg|_{w=1} ,
\end{eqnarray}
describing the linear corrections as well as higher order corrections $c$ and $c_\ast$
introducing a possible correction induced by a non-zero curvature.
Note that the slopes differ from $\rho_{\text{\text{IW}}}$ in the Isgur-Wise limit
introduced in section \ref{New Physics contributions in the IW limit},
since they include the contributions from the coefficient functions $\ffA(w)$
for the $B \to \Ds$ decay and $\ffB^T(w)$ and $\ffB^L(w)$
for the $B \to D$ decay respectively.
Thus, if we evaluate the expansions \eqref{Fexpansion} and \eqref{Gexpansion} up to the first order of magnitude,
the whole information on the right-handed admixtures is contained in the slopes.
Additionally we find the $\rho$ and $\rho_\ast$ to differ from each other,
such that we have the opportunity to calculate a constraint on the right-handed admixture
by comparing the slopes of the two decay modes.
For the $\btoDsln$ decay this implies
\begin{equation}
    \rho_\ast^2 = \rho_{\text{SM}}^2 + \frac{R_1^2(1)}{6} \left[1-\left(\frac{c_+}{c_-}\right)^2\right]\ ,
\end{equation}
where $\rho_{\text{SM}}$ denotes the terms known from the standard model.
In contrast to that the value for $\rho$ concerning $B \to D\,\ell\,\bar\nu$ is left untouched,
since the axial vector component vanishes by parity reasons,
as implied by \eqref{BtoD}--\eqref{BtoDst}.
Therefore we may set $\rho \equiv \rho_{\text{SM}}$ and obtain 
\begin{equation}
    \left(\frac{c_+}{c_-}\right)^2 = 1 - 6\, \frac{\rho_\ast^2-\rho^2}{R_1^2(1)}
\end{equation}
as a measure for the strength of the right-handed admixture.
Using the averaged results 
\begin{eqnarray}
        R_1 = 1.41\pm 0.049 \ &,& \ R_2 = 0.844\pm0.027 \ ,\\
       \rho^2 = 1.18\pm 0.06 \ &,& \ \rho_\ast^2 = 1.24\pm0.04 \ ,
\end{eqnarray}
from the Heavy Flavor Averaging Group \cite{TheHeavyFlavorAveragingGroup:2010qj}
we obtain
\begin{equation}
    \frac{c_+}{c_-} = 0.90\pm 0.09\ ,
\end{equation}
as an estimate for the strength of the right-handed admixture.
Note that we have not included any possible correlations between the errors,
but rather made a naive estimation of the error bars.

Another constraint is given by the fact,
that for the non recoil point $w=1$ the $B \to D\ell\bar\nu$
transition is completely dominated by the vector current,
while in contrast to that the $B\to D^\ast\ell\bar\nu$
transition is proportional to the axial vector current.
Thus, if we include a right-handed admixture,
it is contained in the current experimental results \cite{TheHeavyFlavorAveragingGroup:2010qj},
such that we obtain 
\begin{align}
   c_+ |V_{cb}|\, \ffG(1) &= (42.3\pm1.5)\times 10^{-3}\ ,\\
   c_- |V_{cb}|\, \ffF(1) &= (36.04\pm0.52)\times 10^{-3}\ .
\end{align}
Using lattice data \cite{Bernard:2008dn,Aubin:2004qz,Mackenzie},
\begin{equation}
    \ffG(1) = 1.074 \pm 0.024 , \ \ffF(1) = 0.908 \pm 0.017 ,
\end{equation}
we find
\begin{equation}
    \frac{c_+}{c_-} = 0.99 \pm 0.05\ ,
\end{equation}
while a calculation using the non-lattice values \cite{Uraltsev:2003ye,Gambino:2010bp}
\begin{equation}
    \ffG(1) = 1.02 \pm 0.04 , \ \ffF(1) = 0.86 \pm 0.02 ,
\end{equation}
gives us
\begin{equation}
    \frac{c_+}{c_-} = 0.99 \pm 0.06 ,
\end{equation}
which is in both cases compatible with the standard model value $c_+/c_- \equiv 1$.
Again we have calculated the errors using the assumption,
that no sizable correlations between the experimental measurements and the theoretical values occur.

\section{Summary and Conclusions}
While the left-handedness of the weak interaction is in good agreement
with the data taken from purely left-handed leptonic processes \cite{Michel:1949qe,Bouchiat:1957zz}
the situation for the hadronic interactions is still unclear.
On general grounds one would not expect new physics to show up in a charged current interaction,
but this may as well be a false prejudice.
In this paper we have computed the effect of non-standard couplings for the exclusive semileptonic $\BtoDss$ transition,
which have been introduced in the same way we used for the inclusive semileptonic
$\bar B \to X_c \ell\bar\nu$ decays in \cite{Dassinger:2008as}
using operators of higher dimensions using the most general possible parametrization.
The corresponding dimension six operators then allow new physics effects in charged currents in the hadronic current,
while the leptonic current is left untouched. 

Applying the extended hadronic current including the standard left-handed coupling
as well as the additional right-handed coupling and right- and left-handed vector and scalar couplings
we have calculated the differential decay rates $d\Gamma/dw$ in the Isgur-Wise limit.
Therefore we have introduced new hadronic form factors corresponding to the Dirac-structure of the currents.
The calculation of the differential decay rate also provides us with information about the slopes $\rho$ and $\rho^\ast$
describing the deviation of the differential rate from the zero recoil point at $w = 1$.

The main corrections are the perturbative QCD effects, which are sizable and have to be taken into account.
Due to the vanishing anomalous dimension of the left and the right handed currents the QCD effects are finite for these currents; 
however, additional work is required to compute the virtual corrections to the scalar and tensor currents,
which renormalize under QCD.
Within this paper we have computed the vertex corrections for each occurring current
up to one-loop order including new quark-quark-gluon-boson vertices stemming from the (pseudo)scalar components. Yet the unknown form factors normalizations are still missing and have to be obtained by other methods in order to include these additional structurces, which however are believed to be surpressed.

Comparing the calculated slopes using experimental as well as lattice and non-lattice data
we have been able to calculate bounds on right handed admixtures.
The comparison of the slopes for $B \to D$ and $B \to \Ds$ decays
gives us the result $c_+/c_- = 0.90 \pm 0.09$ using only experimental data.
Furthermore we have used the opportunity to calculate by comparing the experimental results with
lattice and non-lattice data.
Here we obtain $c_+/c_- = 0.99 \pm 0.05$ for lattice
and $c_+/c_- = 0.99 \pm 0.06$ for non-lattice data.
Thus all results are in good agreement with each other and with the purely left-handed standard model current, where $c_+ = c_- = 1$.

\begin{acknowledgments}
We are grateful to Benjamin Dassinger for useful comments. S.T. acknowlegdes helpful discussions with Robert Feger.
This work is supported by the German research foundation DFG under contract MA118/10-1
and by the German Ministry of Research (BMBF),
contracts 05H09PSF.
\end{acknowledgments}

\appendix

\end{document}